\documentclass[prl,twocolumn,aps,10pt,longbibliography,superscriptaddress,preprintnumbers]{revtex4-1} 

\usepackage[T1]{fontenc} 

\usepackage{graphicx}
\usepackage{dcolumn}
\usepackage{bm}
\usepackage{amsmath}
\usepackage{amsfonts}
\usepackage{amssymb}
\usepackage{braket}
\usepackage[separate-uncertainty = true,multi-part-units=single]{siunitx}
\usepackage{subcaption}
\usepackage{color}
\usepackage{hyperref}
\let\oldsqrt\sqrt
\def\sqrt{\mathpalette\DHLhksqrt}
\def\DHLhksqrt#1#2{%
\setbox0=\hbox{$#1\oldsqrt{#2\,}$}\dimen0=\ht0
\advance\dimen0-0.2\ht0
\setbox2=\hbox{\vrule height\ht0 depth -\dimen0}%
{\box0\lower0.4pt\box2}}

\newcommand{\ihat}{\hat{\textbf{\i}}}
\newcommand{\jhat}{\hat{\textbf{\j}}}
\newcommand{\khat}{\hat{\textbf{k}}}

\begin{document}

\title{The viscoelastic signature underpinning polymer deformation under shear flow}

\author{Airidas Korolkovas}
\affiliation{Institut Laue-Langevin, 71 rue des Martyrs, 38000 Grenoble, France}
\affiliation{Department for Physics and Astronomy, L{\"a}gerhyddsv{\"a}gen 1, 752 37 Uppsala, Sweden}
\email{korolkovas@ill.fr}
\author{Sylvain Pr{\'e}vost}
\affiliation{Institut Laue-Langevin, 71 rue des Martyrs, 38000 Grenoble, France}
\author{Maciej Kawecki}
\affiliation{Department for Physics and Astronomy, L{\"a}gerhyddsv{\"a}gen 1, 752 37 Uppsala, Sweden}
\author{Anton Devishvili}
\affiliation{Institut Laue-Langevin, 71 rue des Martyrs, 38000 Grenoble, France}
\affiliation{Department for Physical Chemistry, Naturvetarv{\"a}gen 14, 223 62 Lund, Sweden}
\author{Franz A. Adlmann}
\affiliation{Department for Physics and Astronomy, L{\"a}gerhyddsv{\"a}gen 1, 752 37 Uppsala, Sweden}
\author{Philipp Gutfreund}
\affiliation{Institut Laue-Langevin, 71 rue des Martyrs, 38000 Grenoble, France}
\author{Max Wolff}
\affiliation{Department for Physics and Astronomy, L{\"a}gerhyddsv{\"a}gen 1, 752 37 Uppsala, Sweden}

\date{\today}

\begin{abstract}
Entangled polymers are deformed by a strong shear flow. The shape of the polymer, called the form factor, is measured by small angle neutron scattering. However, the real-space molecular structure is not directly available from the reciprocal-space data, due to the phase problem. Instead, the data has to be fitted with a theoretical model of the molecule. We approximate the unknown structure using piecewise straight segments, from which we derive an analytical form factor. We fit it to our data on a semi-dilute entangled polystyrene solution under \emph{in situ} shear flow. The character of the deformation is shown to lie between that of a single ideal chain (viscous) and a cross-linked network (elastic rubber). Furthermore, we use the fitted structure to estimate the mechanical stress, and find a fairly good agreement with rheology literature.
\end{abstract}


\maketitle

\section{Introduction}
\begin{figure}[ptbh!] 
		\begingroup
			\sbox0{\includegraphics{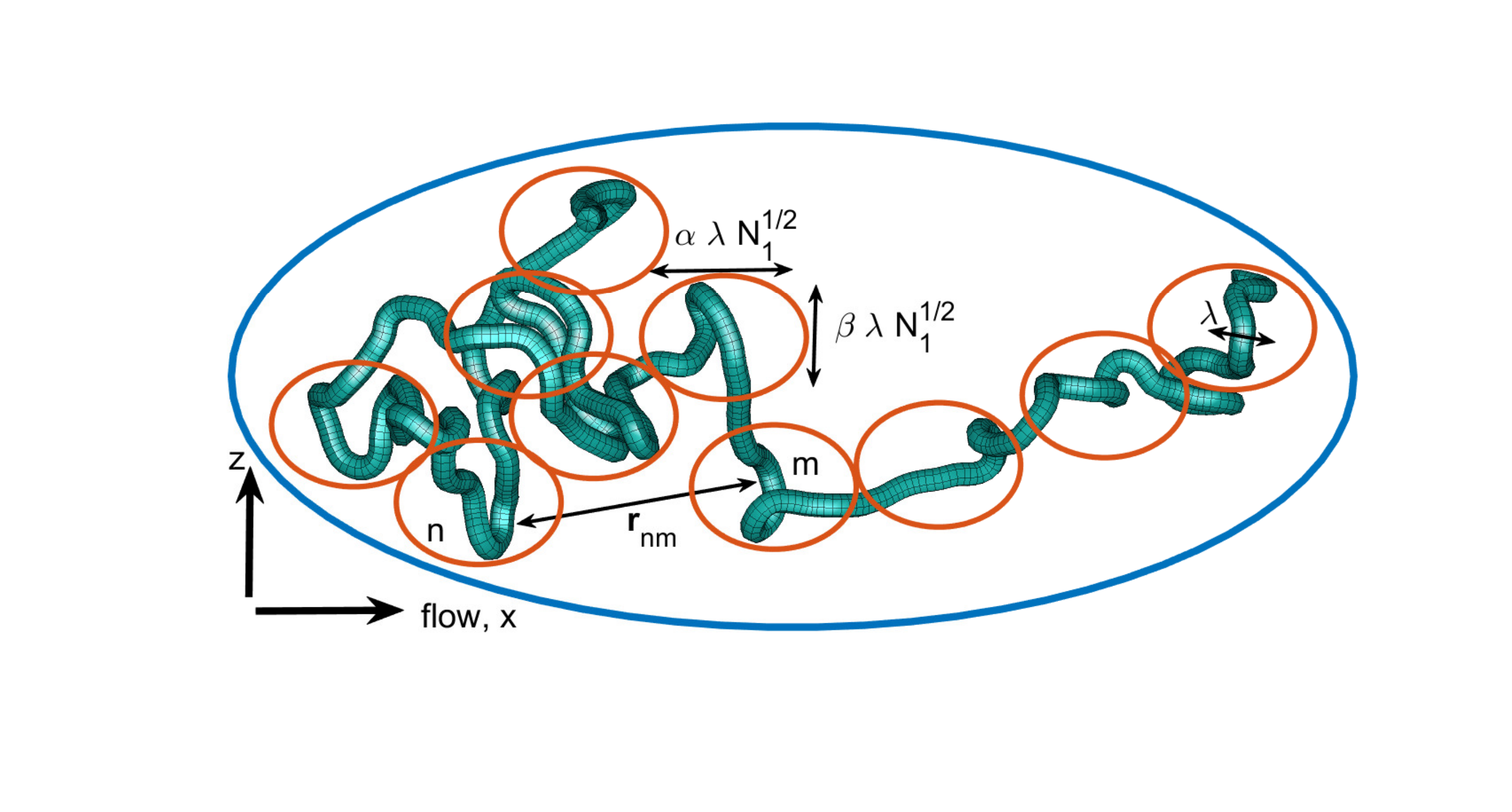}} 
			\includegraphics[clip,trim={.1\wd0} {.1\wd0} {.1\wd0} {.05\wd0},width=\linewidth]{./fig/multiscale2.pdf} 
		\endgroup
\caption{An entangled polymer chain under shear flow. Its structure is defined by the distribution of distances between all monomer pairs $(n,m)$. Conceptually, and for the purpose of data analysis, it is subdivided into two (or more) linear regimes, $\braket{\mathbf{r}_{nm}^2} \propto |n-m|$, each an ideal random walk. On the short scale (red ellipses), the anisotropy is just a few percent, but grows much bigger on a large scale (blue ellipse). The chain gradually stretches along the flow, and shrinks perpendicular to it.}\label{multiscale}
\end{figure}

Viscoelastic materials have properties of both viscous liquids and elastic solids. Such non-Newtonian fluids are very common, from daily items like food and cosmetics, to raw materials for plastics and fibres. Their complex response to flow is due to the intricate deformation of polymer molecules, shown in Fig.~\ref{multiscale}. To describe it, let us consider two extreme cases. On one hand there are fully elastic materials, like rubber, which are composed of permanently cross-linked polymer chains. Under stress they exhibit the so-called affine deformation, meaning that the mean square distance (MSD) between two monomers $n$ and $m$ is linearly proportional to their separation $\braket{\mathbf{r}_{nm}^2} \propto |n-m|$. On the opposite side, there is the ideal chain (Rouse model), whose deformation is non-affine, scaling as $|n-m|^2$, see Ref.~\cite{pincus1977dynamics}. In this article we examine an intermediate case, a semi-dilute entangled polystyrene solution, and using a novel fitting approach we show that the deformation is proportional to $|n-m|^\xi$, where $1<(\xi=1.2)<2$ is a viscoelastic signature exponent.

Thanks to deuteration, small angle neutron scattering (SANS) can measure the structure of an individual polymer chain, called the form factor. Many previous studies have used extensional flow to characterize the relaxation of polymers (creep) over time~\cite{muller1990polymer, lopez2017chain, wang2017fingerprinting}. In this work we focus on shear flow, which poses more practical challenges, but has an advantage of eschewing the complicated time response, once steady state has been reached. The effect of a shear rate $\kappa$ on the material structure is quantified with a dimensionless Weissenberg number $\text{Wi} = \kappa \tau$, where $\tau$ is the relaxation time specific to each fluid, typically a millisecond or more. For polymer melts~\cite{wignall1981measurements}, shear can be applied on a heated sample, which is then quenched below the glass transition temperature, and the molecular structure is later examined \emph{ex situ}. Polystyrene (PS) has been measured with SANS using this technique, at an estimated shear rate of $\text{Wi} = 4$. An asymmetry of 1.7 was detected between the chain radii of gyration along the flow and the vorticity directions~\cite{Muller1993}. More difficult, but also more industrially relevant experiments measure the fluid structure under \emph{in situ} shear. Molten polymers like polydimethylsiloxane (PDMS) and polybutadien (PBD) are popular examples, thanks to their low glass transition temperature and comparatively low viscosity. \emph{In situ} steady flow SANS experiments have not detected any anisotropy of the form factor for either of these samples. The highest shear rate for the PBD experiment~\cite{Noirez2009} in Couette geometry was $\text{Wi} = 5.4$ and for the PDMS experiment~\cite{kawecki2018direct} in cone-plate geometry it was $\text{Wi} = 0.8$. The only \emph{in situ} shear experiment that has shown anisotropy of entangled polymers was performed in a Couette cell with PS at $\text{Wi}\approx 1$, but since the relaxation time has not been reported, the Weissenberg number is uncertain~\cite{Yearley2010}. Anisotropy of 1.5 has also been detected in a dilute solution of long but unentangled PS chains~\cite{lindner1988shear}.

Up to now, the form factor of entangled semi-dilute polymer solutions has not been characterized by SANS under \emph{in situ} shear. The advantage of the semi-dilute condition is that it has a lower viscosity and glass transition temperature than a melt, facilitating its handling and enabling higher Wi. However, shear induces massive concentration fluctuations, leading up to complete demixing in the extreme case. The resulting SANS signal contains a strong contribution from the structure factor~\cite{nakatani1994neutron, morfin1999temperature}, hindering the single chain analysis~\cite{Nakatani1994}. Fortunately, it is possible to use deuteration to match the contrast between the solvent and the polymer, fully canceling the inter-chain contribution to scattering, even at high density~\cite{Hammouda2008}.

The scenarios where polymers may deform range from dilute, to semi-dilute, to melts. Moreover, a strong anisotropy can also be found in cross-linked polymer networks of gels and rubbers~\cite{read1997lozenge,basu2011nonaffine}, as well as nanocomposites like polymer-clay~\cite{takeda2010rheo,schmidt2002small}. Mechanically, these materials are probed in either shear or extensional flow, applied in a steady, oscillatory, or stepwise mode, or even a superposition of multiple stimuli. As a rule of thumb, a strong deformation will stretch the polymer along flow and shrink it perpendicular to flow. However, the detailed shape of the form factor can have considerable differences in the various cases listed above. While many isotropic theories exist for equilibrium~\cite{kholodenko1993analytical, pedersen1996scattering, sigel2018form}, anisotropic scattering patterns up to now have been analyzed in mostly \emph{ad hoc} fashion: fitting 1D radial cuts~\cite{Muller1993}, comparing angular sector averages~\cite{Yearley2010}, fitting ellipses to isointensity curves~\cite{Muller1993}, and fingerprinting with spherical harmonics~\cite{wang2017fingerprinting}. In the present work, we develop a new approach to extract the underlying real-space structure directly from the data, not requiring any knowledge of the molecular motion. The observed form factor originates from the MSD between the monomers, which is a function of their index separation along the chain, see Fig.~\ref{multiscale}. At equilibrium, this function is a straight line (ideal random walk), while under a strong deformation it becomes some other, unknown curve. Our main novelty is to approximate this curve with a set of straight segments, or layers. This discrete model converges to the exact mathematical result when the number of segments is brought to infinity (a textbook definition of the Riemann integral). Luckily, in real-world experiments the MSD deviation from a perfect straight line is quite small, almost never exceeding $\times 2$, so there is no need for an infinity of parameters for a good description, and only a few layers are sufficient. In this case, the model is convenient to integrate analytically, and the resulting formula is fitted to the 2D data, to determine the width and the slope of each layer. This structure is then fitted to reveal a power-law of $\xi=1.2$, and that is our novel measure of structural non-affinity.

\section{Experimental}
An unlabeled polymer solution of $C$ chains with $N$ monomers each is characterized by a quantity known as the structure factor
\begin{equation}\label{Sdef}
S(\mathbf{q}) = \frac{1}{NC} \Big| \sum_{n=1}^{N} \sum_{c=1}^{C} e^{-i\mathbf{q}\cdot \mathbf{r}_{nc}}\Big|^2 = \frac{|F|^2}{NC}
\end{equation}
which is the modulus squared of the Fourier transform $F$ of all the monomer positions $\mathbf{r}_{nc}$. While there are $(NC)^2$ terms in the double sum, only the nearest neighbours of each scatterer contribute to the structure, hence it is normalized by $NC$. With this convention, a structureless fluid (i.e. ideal gas) has $S(\mathbf{q}) = 1$. In real fluids, one can measure deviations from this baseline which is a signature of their molecular interactions~\cite{pedersen2004scattering}. However, the focus in this study is to obtain the single chain form factor, defined as
\begin{equation}\label{Pdef}
P(\mathbf{q}) = \frac{1}{N^2} \Big|\sum_{n=1}^N e^{-i\mathbf{q}\cdot \mathbf{r}_n}\Big|^2
\end{equation}
with the normalization chosen to have $P(\mathbf{q}\rightarrow 0) = 1$. Using a mixture of deuterated (D, phase~1) and hydrogenated (H, phase~2) chains, it is possible to isolate the form factor $P(\mathbf{q})$ even in dense solutions where the chains strongly overlap. This method, called the Zero Average Contrast, is described in the handbook Ref.~\cite{Hammouda2008} (see Eq.~35 on page 324), and is a standard SANS technique. Here we briefly outline its derivation. The experimental scattering cross-section from a sample of volume $V$ consists of three terms:
\begin{equation}\label{Vsigma}
\frac{d\Sigma}{d\Omega} = \frac{1}{V}|b_S F_0 + b_D F_1 + b_H F_2|^2
\end{equation}
where $F_{0,1,2}$ are the Fourier transforms of the solvent, D, and H monomer positions respectively, while $b_{S,D,H}$ are the corresponding scattering lengths of each nuclear species. While the volume $v$ of one monomer and $v_S$ of one solvent molecule are in general different, the system can be assumed to be incompressible, leading to: $v_S F_0 + v F = 0$, where $F=F_1+F_2$ is the Fourier transform of all polymers as defined in Eq.~\eqref{Sdef}. The solvent term $F_0$ is plugged into Eq.~\eqref{Vsigma}, leaving only the polymer part:
\begin{multline}\label{Ftrans}
\left(\frac{V}{v^2}\right)\frac{d\Sigma}{d\Omega} = |\rho_1 F_1 + \rho_2 F_2|^2 = \\
\rho_1 \rho_2 |F|^2 + \rho_1(\rho_1-\rho_2) |F_1|^2 + \rho_2 (\rho_2-\rho_1) |F_2|^2
\end{multline}
For convenience, the scattering length density (SLD) contrast has been defined as $\rho_1 = b_D/v - b_S/v_S$ and $\rho_2 = b_H/v - b_S/v_S$ for the two labels. The Fourier transform squared of each phase can be further decomposed into the diagonal (intra-chain) and the off-diagonal (inter-chain) terms:
\begin{equation}\label{F1}
|F_1|^2 = C_1 N^2 P(\mathbf{q}) + C_1(C_1-1)Q(\mathbf{q}),
\end{equation}
and similarly for $F_2$ and $F$. Note that the total number of chains $C_1+C_2=C$ is fixed. The auxiliary function
\begin{subequations}
\begin{align}
Q(\mathbf{q}) &= \sum_{n,m=1}^N \braket{e^{-i\mathbf{q}\cdot \mathbf{r}_{\alpha n, \beta m}}}_{\alpha \neq \beta}\\
&= \frac{NS(\mathbf{q})-N^2 P(\mathbf{q})}{C-1}\label{Qaux}
\end{align}
\end{subequations}
is the interference between any two different chains $\alpha \neq \beta$. The definition of $Q(\mathbf{q})$ involves only the monomer positions, not their SLD, since the contrast information has already been factored out in Eq.~\eqref{Ftrans}. The weight in front of $Q(\mathbf{q})$ is proportional to $C^2$, whereas the weight of $P(\mathbf{q})$ has a $C^1$ dependence, and this difference enables the tuning of the relative contributions of the form and the structure factors. Eq.~\eqref{Qaux} is plugged into Eq.~\eqref{F1}, which is then plugged into Eq.~\eqref{Ftrans}, revealing the scattered intensity in terms of the form and the structure factors only:
\begin{multline}
\left(\frac{\phi_1+\phi_2}{v}\right) \frac{d\Sigma}{d\Omega} =\\ (\rho_1-\rho_2)^2 \phi_1 \phi_2 N P(\mathbf{q}) + (\phi_1 \rho_1 + \phi_2 \rho_2)^2 S(\mathbf{q})
\end{multline}
It is the same formula as used in other SANS studies~\cite{hammouda2015single}. In particular, it shows that if we set the average contrast to $\phi_1 \rho_1 + \phi_2 \rho_2 = 0$, the structure factor contribution $S(\mathbf{q})$ vanishes, since the three inter-chain signals from hPS-hPS, dPS-dPS, and hPS-dPS add up to zero in this case.

\begin{figure}[ptbh!]
		\begingroup
			\sbox0{\includegraphics{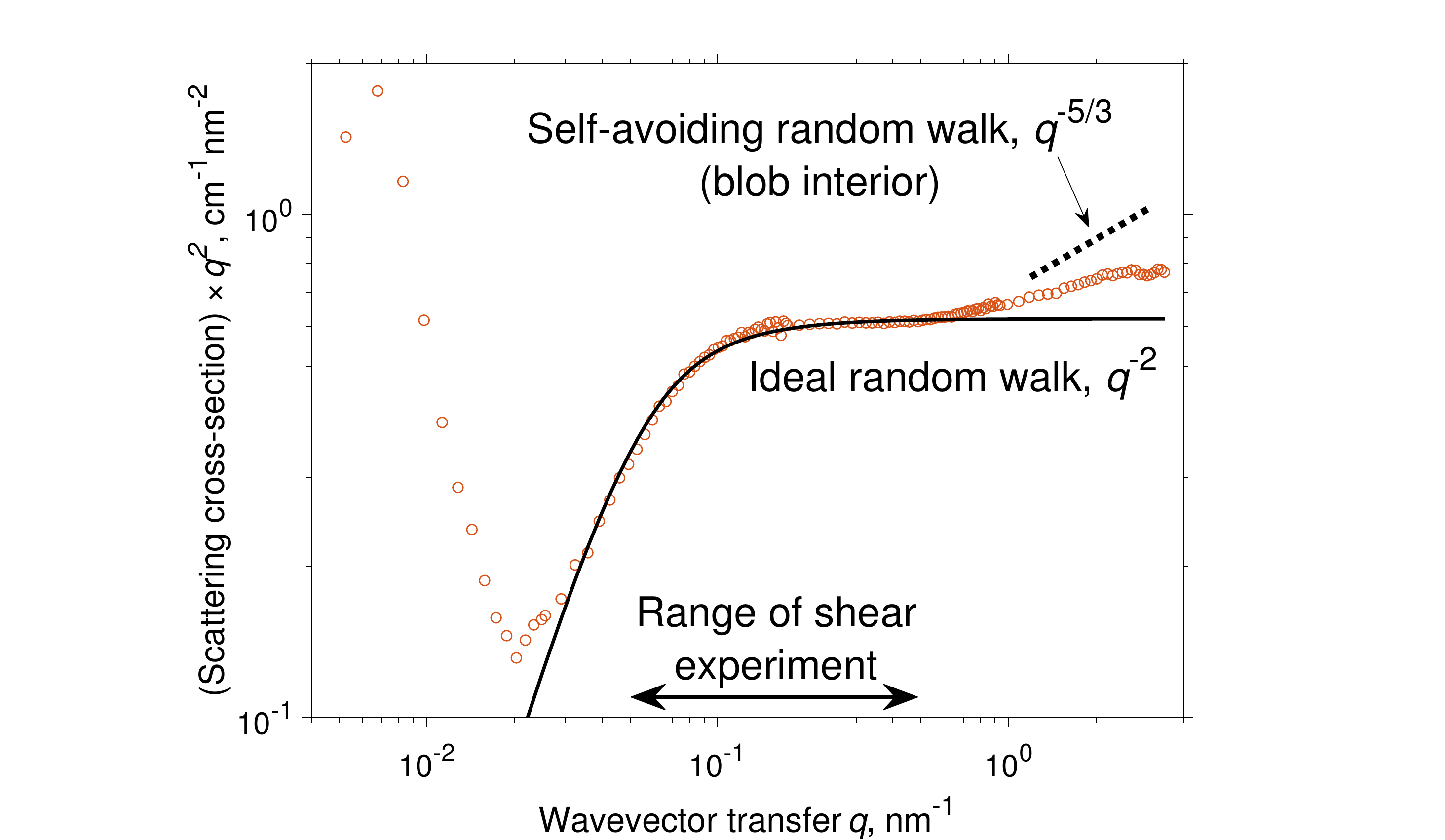}} 
			\includegraphics[clip,trim={.1\wd0} 0 {.15\wd0} 0,width=\linewidth]{./fig/static.pdf} 
		\endgroup
\caption{The scattering cross-section from a quiescent solution, multiplied by $q^2$ to reveal the ideal random walk character (flat line) of the chain form factor. The solid black line is the Debye function, Eq.~\eqref{debye}, fitted to $R = \SI{27.12}{\nano\meter}$.}\label{static}
\end{figure}

Our sample was an entangled semi-dilute polymer solution, with a volume fraction $\phi_1 = 0.172$ of deuterated PS (\SI{575}{\kilo\gram\per\mole}, $N=5127$, $\text{PDI} = 1.09$) and $\phi_2 = 0.0998$ of hydrogenated PS (\SI{510}{\kilo\gram\per\mole}, $N=5000$, $\text{PDI} = 1.1$), purchased from Polymer Source. It was prepared by first dissolving the powdered PS mix in a glass beaker with a large amount of deuterated toluene, using a magnetic stirrer. After removing the stirrer, the solution was left in a ventilated fume hood for several days until the toluene has evaporated to the volume fraction quoted above, which was determined by weighing the dry and the dissolved polymer, minus the container. The detailed rheological characterization of a similar sample has been reported in Ref.~\cite{korolkovas2017polymer}. 

In our region of interest, $\mathcal{O}(qR) = 1$, the structure and the form factors as defined in Eqs.~\eqref{Sdef} and \eqref{Pdef} both have a similar magnitude of $\mathcal{O}(S(q)\approx P(q)) = 1$. Using the SLD values $\rho_1 = \SI{0.47e10}{\centi\meter^{-2}}$ and $\rho_2 = \SI{-4.5e10}{\centi\meter^{-2}}$ we can estimate the ratio of the two intensities as
\begin{equation}
\frac{ (\phi_1 \rho_1 + \phi_2 \rho_2)^2}{(\rho_1-\rho_2)^2 \phi_1 \phi_2 N} = \SI{6e-5}{} \ll 1
\end{equation}
which is quite small, thanks also to the high degree of polymerization $N=5000$. Even though our system is not exactly contrast-matched, the structure factor contribution is negligible beyond $q>\SI{0.04}{\nano\meter^{-1}}$, see Fig.~\ref{static}. This quiescent data~\cite{d11data} was recorded on the instrument D11 (Institut Laue-Langevin, Grenoble, France) providing a wide $q$ range, in this case \SIrange{0.004}{4}{\per\nano\meter}, covering distances from multi-chain clusters to a single monomer. Our focus is on intermediate $q$ values, which are well fitted with the Debye function, Eq.~\eqref{debye}, establishing the radius of gyration to be $R=\SI{27.12}{\nano\meter}$. To assess the validity of this fit, we compare it with literature data~\cite{fetters1994molecular} for dilute PS of the same molecular weight in toluene $R_{\text{TOL}} = \SI{29.9}{\nano\meter}$ (maximum swelling in a good solvent), and in cyclohexane $R_{\text{CH}} = \SI{20.2}{\nano\meter}$ (a theta solvent which fully screens the excluded volume). Our semi-dilute solution of volume fraction $\phi=\phi_1+\phi_2=0.27$ is partially screened, so the radius is estimated to be $(1-\phi) R_{\text{TOL}} + \phi R_{\text{CH}} = \SI{27.3}{\nano\meter}$, in agreement with our data.

The form factor of an ideal random walk has a power law behaviour of $P \propto (qR)^0$ for $qR\ll 1$ and $P \propto (qR)^{-2}$ for $qR\gg 1$, as evidenced in Fig.~\ref{static}. Eventually at high $q$ the scattering starts to probe correlations inside the blob of size $\mathcal{O}(\lambda) = \SI{1}{\nano\meter}$, which is a typical distance between the semi-dilute chains, called the mesh size. Within the blob ($q\lambda \gg 1$) the excluded volume interactions are not screened, so the polymer form factor changes towards the scaling of $P \propto (q\lambda)^{-5/3}$, which is a signature of a self-avoiding random walk (see textbook Refs.~\cite{deGennesScalingConcepts, doi1988theory}). In addition, the scattering from density fluctuations at the chemical monomer level may become visible for the highest $q$-values (not measured here). On the opposite side of the spectrum, the ultra low $q$ data also deviates from Debye, this time due to scattering from very slowly relaxing density inhomogeneities spanning large distances, likely hundreds of chains or more~\cite{morfin1999temperature}. Extreme viscoelastic samples like ours are difficult to fully equilibrate, as some residual flow persists for many hours if not days (one experiment has been running for almost 100 years~\cite{edgeworth1984pitch}). Even when left perfectly still, the sample may keep flowing due to an interplay of gravity and the capillary forces between the narrow gap of the rheometer plates. This can induce concentration fluctuations (see Refs.~\cite{wu1991enhanced, hashimoto1992butterfly, groisman2000elastic, saito2002structures}), and while their amplitude may be tiny, when integrated over a long distance, a strong SANS signal can result at ultra low $q$.

Our shear experiments were conducted on PAXY (Laboratoire L{\'e}on Brillouin, Saclay, France), with a narrower $q$ range set at \SIrange{0.05}{0.5}{\per\nano\meter}, where the scattering is fully described by the Debye function. We have used a custom-made vertical sealed cone-plate shear cell~\cite{Kawecki2016}, designed for both SANS and NSE instruments and allowing a smaller liquid volume than typical Couette cells~\cite{Yearley2010}, which can be a considerable advantage for costly and rare deuterated samples. It is also well suited for shearing fluids which exhibit non-linear viscoelastic phenomena such as the rod-climbing effect. A vertical cone-plate geometry is a necessity for rheo-NSE and allows a direct measurement of both structure (SANS) and dynamics (NSE) in the same setup. In our experiment the shear rate was $\kappa = \SI{300}{\per\second}$, corresponding to $\text{Wi}=30$. Data collection has lasted \SI{4.25}{\hour} per spectrum, at a temperature of \SI{45}{\celsius}, which is the same as used at D11 for the quiescent measurement.

In this article we only report data from the SANS experiment carried over one day. After that, the experiment continued for three more days with NSE, which will be a separate subject. However, for full disclosure we note that after these four days of shearing, we have spotted some wear of the cell sealing, causing aluminum and teflon impurities to have leached into the sample. Solid particles are known to give rise to Porod scattering of $P\propto q^{-4}$, and fortunately there was no trace of it in the range covered by PAXY, where the Debye law $P\propto q^{-2}$ dominates. As the SANS data was collected during the first day of shearing, the impurities at that stage must have been very dilute and hence invisible to the beam. On top of that, the particle size must have been much greater than the polymer radius of gyration, falling outside of the SANS range. Such big particles cannot interfere with the polymer dynamics, as that is only possible in polymer-nanocomposites where the two components are similar-sized~\cite{schmidt2002small}. These specialty materials require advanced chemical synthesis and cannot be produced by just using mechanical friction to grind up some aluminum dust. Therefore, even if we would have had a considerable percentage of sample contamination, its effect could not have altered the entanglement physics, but only lowered the overall polymer density. This means that the actual Wi may have been 29 instead of 30 we claim. Either way, it is unlikely that these impurities could have altered the polymer form factor beyond the uncertainty of the fit (\SI{15}{\percent}), as explained in the next section.

\section{Results}
\begin{figure*}[ptbh!] 
\begin{subfigure}{.49\textwidth}
\begingroup
			\sbox0{\includegraphics{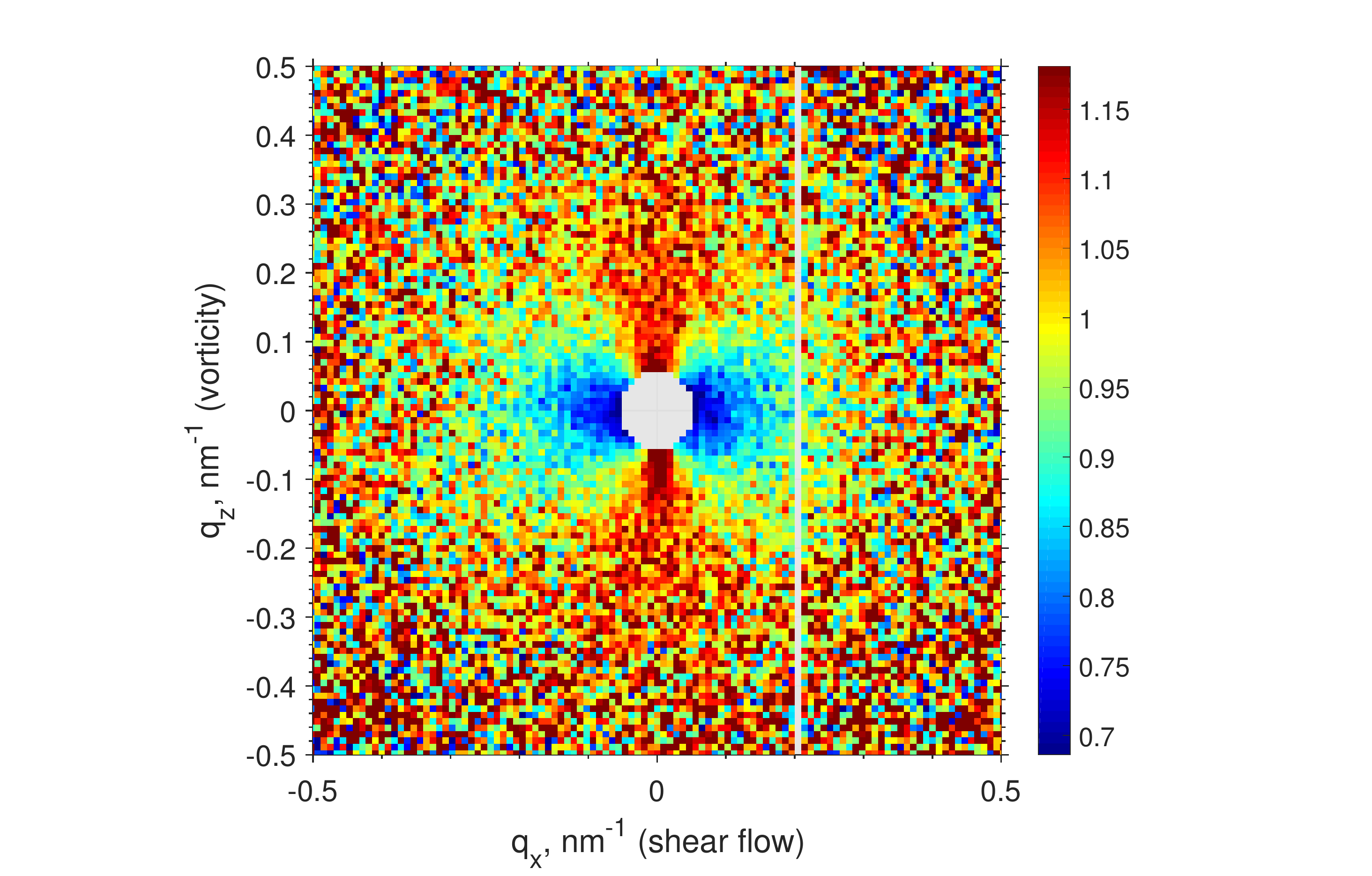}} 
			\includegraphics[clip,trim={.13\wd0} 0 {.17\wd0} 0,width=\linewidth]{./fig/hiQexp.pdf} 
		\endgroup\caption{Experimental data}\label{hiQexp}
\end{subfigure}
\hfill
\begin{subfigure}{.49\textwidth}
		\begingroup
			\sbox0{\includegraphics{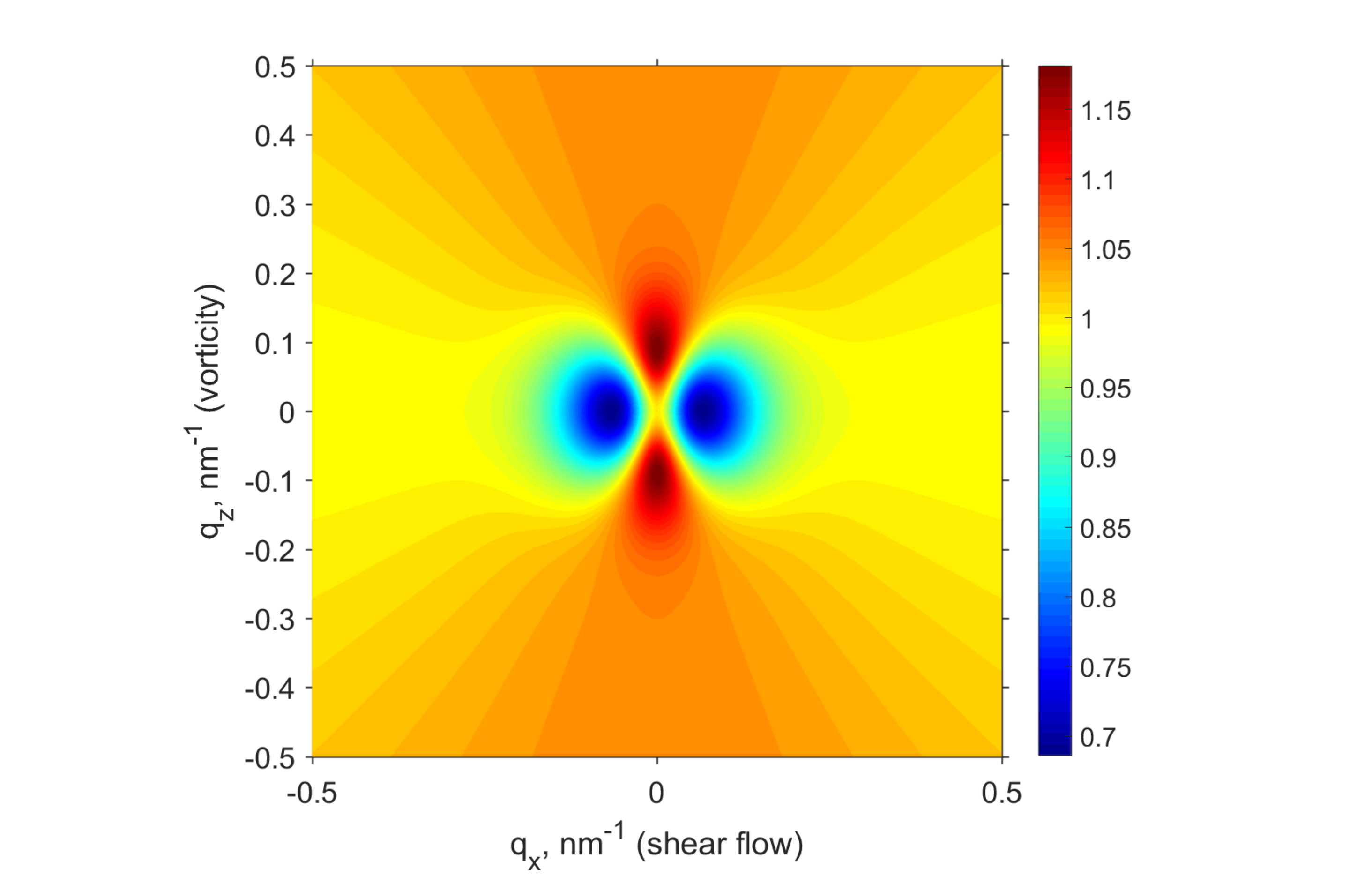}} 
			\includegraphics[clip,trim={.13\wd0} 0 {.17\wd0} 0,width=\linewidth]{./fig/hiQfit.pdf} 
		\endgroup\caption{Analytical fit, Eq.~\eqref{fullq}}\label{hiQfit}
\end{subfigure}

\begin{subfigure}{.49\textwidth}
		\begingroup
			\sbox0{\includegraphics{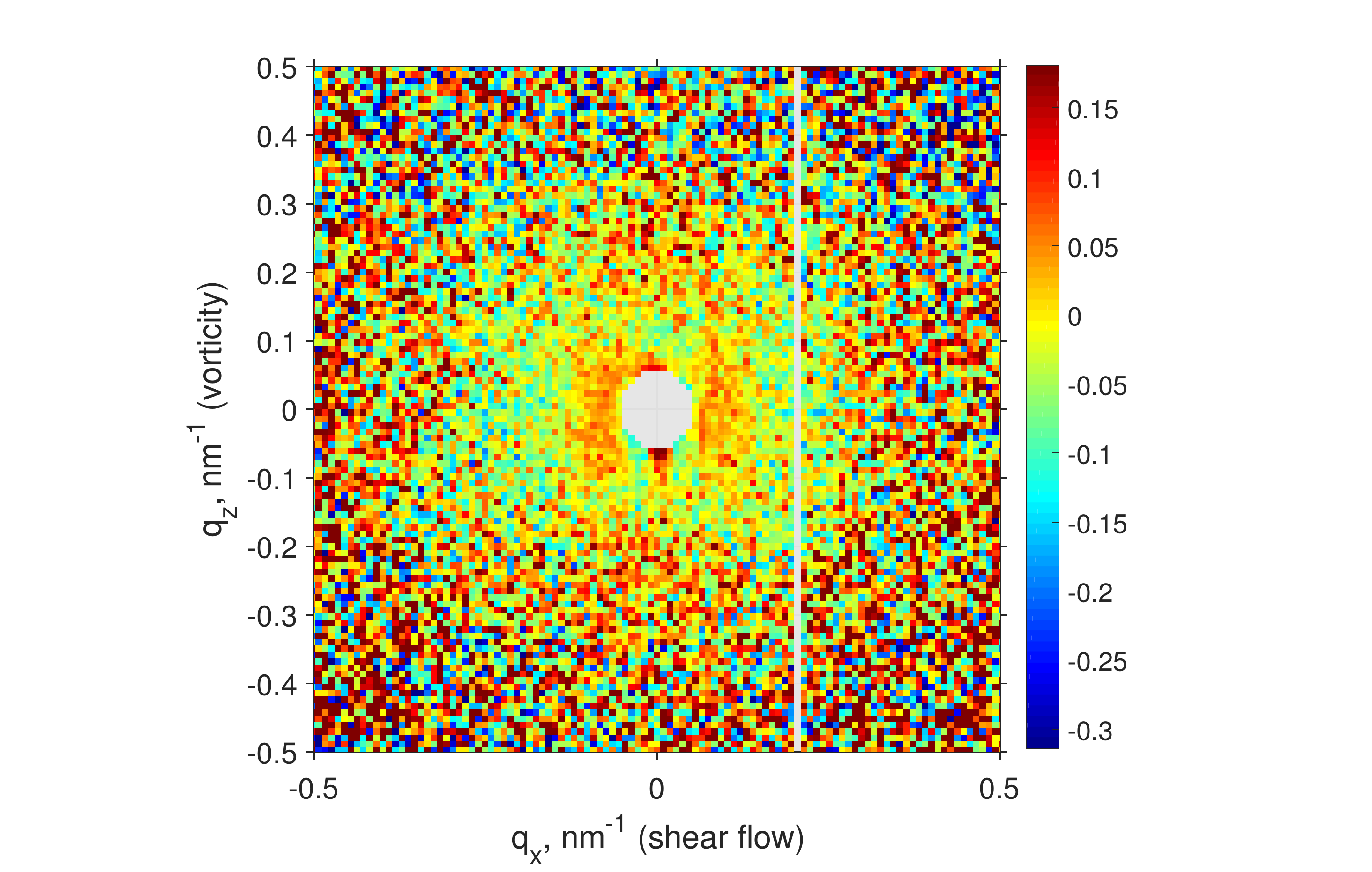}} 
			\includegraphics[clip,trim={.13\wd0} 0 {.17\wd0} 0,width=\linewidth]{./fig/residuals.pdf} 
		\endgroup\caption{Residuals plot}\label{residuals}
\end{subfigure}
\hfill
\begin{subfigure}{.49\textwidth}
		\begingroup
			\sbox0{\includegraphics{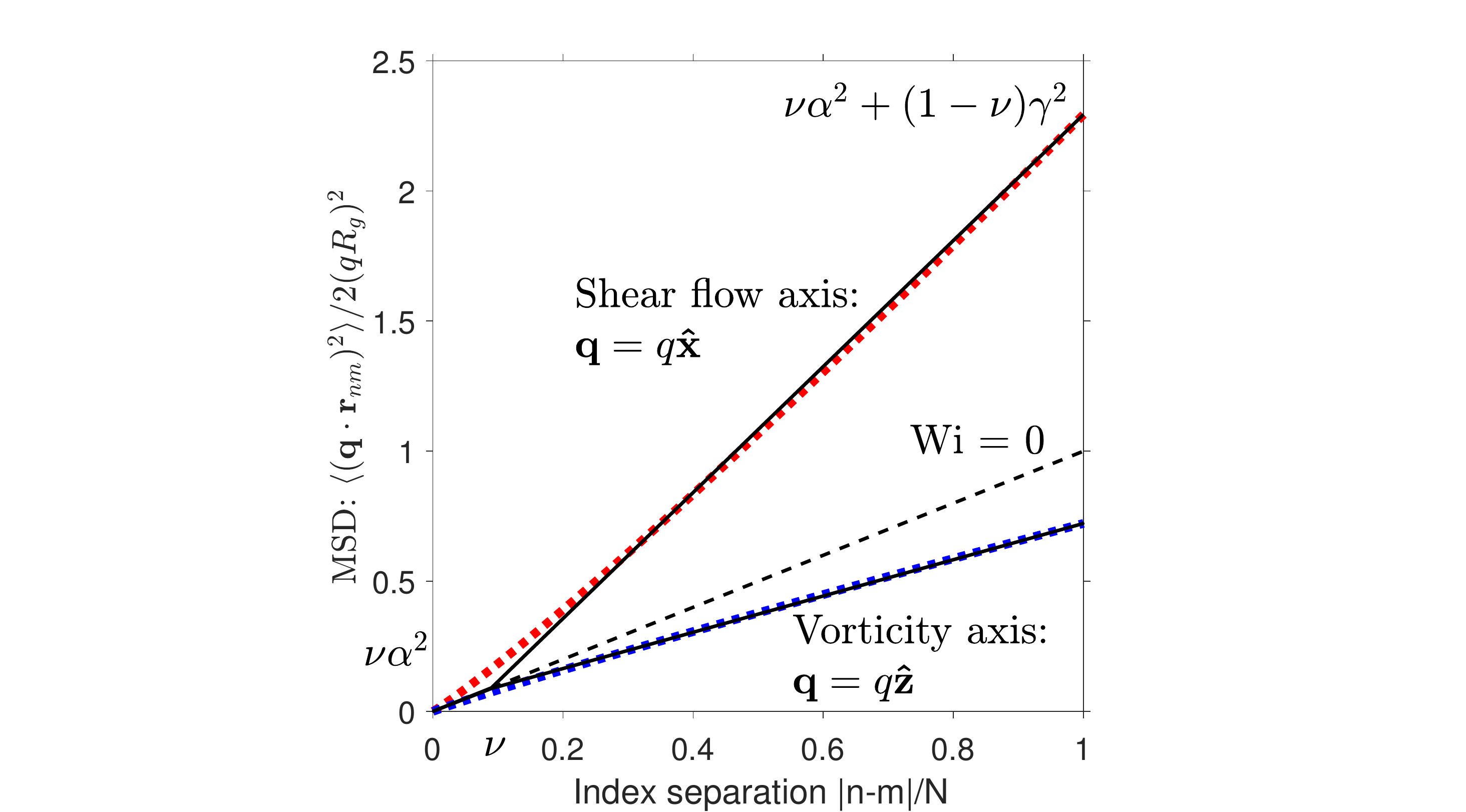}} 
			\includegraphics[clip,trim={.2\wd0} 0 {.2\wd0} 0,width=\linewidth]{./fig/schemeplot.pdf} 
		\endgroup\caption{Real-space structure}\label{model}
		\end{subfigure}
\caption{(a-c) The scattered intensity under shear $P(\mathbf{q})$, divided by the quiescent signal $P_{\text{iso}}(\mathbf{q})$. This plot removes the Debye envelope $1/q^2$, highlighting the structural changes induced by the shear. The data (a) is fitted with the analytical function (b), and their difference is plotted in (c), showing that the fit accounts for \SI{85}{\percent} of the signal or more. (d)~The inferred mean square distance (MSD) between two monomers, for different directions probed by the scattering vector $\mathbf{q}$. The dotted lines are a sketch of what the true function may look like. The straight black lines show our piecewise approximation, Eq.~\eqref{regimes}. The dashed black line is the isotropic MSD found at equilibrium, Eq.~\eqref{iso}.}\label{exp}
\end{figure*}
\subsection{Theory}
Previous studies on sheared polymers have for the most part focused on the deformation as a function of shear rate, or time in the case of extensional flow~\cite{wang2017fingerprinting}. In the present experiment we elucidate the lesser understood aspect, which is the structural, or the $q$ dependence. For this goal, the entire available beamtime was devoted to only one shear rate $\text{Wi} = 30$, the highest possible with the setup. In Fig.~\ref{hiQexp} we plot the 2D scattering pattern under shear, divided by the quiescent signal. Anisotropy at low $q$ is clearly visible, showing that the chains stretch along the flow and shrink along the vorticity, but the extent of the deformation decreases as we probe deeper into the chain interior (higher $q$). The observed signal originates from the distribution function $\Psi(\mathbf{r}_{nm})$ of the distance $\mathbf{r}_{nm} = x\ihat + y\jhat + z\khat$ between the monomers $n$ and $m$ (we drop the subscript $nm$ from now on). In equilibrium, it is well described by a Gaussian function (the normalization factor is not shown):
\begin{equation}
\Psi(\mathbf{r}, \text{Wi}\ll 1) = \exp \left(-\frac{x^2+y^2+z^2}{2|n-m|\lambda^2}\right)
\end{equation}
This result is exact for an infinite ideal random walk, and is very often applied for real polymers too. Under an experimentally reasonable amount of deformation, our assumption is that the functional form of the distribution $\Psi$ remains close to a Gaussian, but its shape is now a tilted ellipsoid rather than a sphere:
\begin{multline}\label{anisogauss}
\Psi(\mathbf{r}, \text{Wi}\gtrsim 1) =\\
\exp \left(-\frac{1}{2}\left(A_{xx}x^2 + 2A_{xy}xy + A_{yy}y^2+ A_{zz}z^2\right)\right)
\end{multline}
The ellipse is specified by the anisotropy matrix $A_{ij}(|n-m|)$. In other words, we account for the observed deformation of the polymer form factor through a change of the Gaussian's dimensions and orientation, rather than a change of the function itself. It is justified, since the scattering is mainly sensitive to the width of the monomer distribution, while its precise shape is less important. Nevertheless, for extreme deformations this assumption may break down, in which case we could extend Eq.~\eqref{anisogauss}, for example by adding higher order terms of the Hermite expansion. This would introduce additional fitting parameters, which would enable an experimental determination of the magnitude of those extra terms. For now, only the zeroth order term (a regular Gaussian) is considered, as it will be seen to already produce a satisfactory fit to the data. In this case, the scattering contribution from two monomers is (see Appendix~2.1 in Ref.~\cite{doi1988theory}):
\begin{equation}\label{expi}
\braket{e^{i\mathbf{q}\cdot \mathbf{r}}} = e^{-\braket{(\mathbf{q}\cdot \mathbf{r})^2}/2}
\end{equation}
The argument of the exponent, which we call the MSD function,
\begin{equation}\label{msddef}
\braket{(\mathbf{q}\cdot \mathbf{r})^2} = \braket{x^2} q_x^2 + 2\braket{xy} q_x q_y + \braket{y^2} q_y^2 + \braket{z^2} q_z^2
\end{equation}
contains 4 averages, derived from the 4 components of the anisotropy matrix $A_{ij}$:
\begin{equation}\label{averages}
\begin{pmatrix}
\braket{x^2}\\ \braket{y^2}\\ \braket{xy}\\ \braket{z^2} \end{pmatrix} = 
\begin{pmatrix} D/A_{xx}\\ D/A_{yy}\\ (1-D)/A_{xy}\\ 1/A_{zz}\end{pmatrix}
\end{equation}
where we have defined $D = 1/\left(1-A_{xy}^2/(A_{xx}A_{yy})\right)$ for brevity. We now plug in Eq.~\eqref{expi} to Eq.~\eqref{Pdef}, which in the continuous limit becomes a double integral
\begin{equation}\label{Sq}
P(\mathbf{q}) = \frac{1}{N^2}\int_0^N dn \int_0^N dm\, \exp \left(-\frac{\braket{(\mathbf{q}\cdot \mathbf{r}_{nm})^2}}{2} \right)
\end{equation}
An exact analytical solution is available in equilibrium (see Section~2.4 in Ref.~\cite{doi1988theory}), since the argument
\begin{equation}\label{iso}
\braket{(\mathbf{q}\cdot \mathbf{r}_{nm})^2}/2 = a|n-m|/N
\end{equation}
is then a straight line with a constant dimensionless slope $a = (qR)^2$, where $R^2=N\lambda^2/6 = (\SI{27.12}{\nano\meter})^2$ is the equilibrium radius of gyration. The result is known as the Debye function:
\begin{equation}\label{debye}
P_{\text{iso}}(\mathbf{q}) = \frac{2\left( e^{-a}-1+a\right)}{a^2}
\end{equation}
Under shear, the slope is not constant, as evidenced by the $q$-dependence of the anisotropy, Fig.~\ref{hiQexp}. The functional form of the MSD dependence on $|n-m|$ is due to the specific molecular and topological interactions, and at present no suitable theory is available for entangled polymer solutions. Luckily, this information is not necessary to fit the SANS data, and we propose to integrate Eq.~\eqref{Sq} by approximating the unknown MSD function with a series of straight lines. Any reasonable curve can be approximated to an arbitrarily high accuracy with a set of ever shorter segments. In this study we only use two of them:
\begin{equation}\label{regimes}
\frac{\braket{(\mathbf{q}\cdot \mathbf{r}_{nm})^2}}{2} = \begin{cases}
b|n-m|/N, & |n-m|< N_1\\
c|n-m|/N + (b-c)\nu, & |n-m|> N_1
\end{cases}
\end{equation}
Here $N_1$ is the first layer ``thickness'', specified through the fitting parameter $0<(\nu=N_1/N)<1$. In our cone-plate experiment, only the $xz$ plane could be measured, so the slopes from Eq.~\eqref{msddef} reduce to
\begin{align}
b &= R^2\left[(\alpha q_x)^2 + (\beta q_z)^2\right]\\
c &= R^2\left[(\gamma q_x)^2 + (\delta q_z)^2\right]
\end{align}
although if 3D data was available from Couette or \emph{ex situ} experiments, the full Eq.~\eqref{msddef} would be retained. Now the two slopes in two dimensions are described by four fitting parameters $(\alpha,\beta,\gamma,\delta)$, which trace back to the original inter-monomer distribution, Eq.~\eqref{anisogauss}. Our generic model is plugged into the scattering function, Eq.~\eqref{Sq}, and integrated piece by piece to yield:
\begin{multline}\label{fullq}
P(\mathbf{q}) = \frac{2}{b^2} \Bigl( b-1 + e^{-\nu b} \times \\
\left\{1 + [(b/c)^2-1]e^{(\nu-1)c} + (1-\nu)(b/c)(b-c)\right\} \Bigr)
\end{multline}
which is our main result. In principle, more than two layers can be added, keeping in mind that every new layer introduces three fitting parameters (its thickness and the two $(x,z)$ slopes). Although the formulas become tedious with extra layers, the analytical solution always exists and is straightforward to obtain with symbolic algebra software (Matlab, Mathematica, etc.). 

\subsection{Experimental application}
\begin{table}[htb]
\begin{tabular}{l | l l l}
										&	Slope $x$	& Slope $z$	& Thickness\\
										\hline
Layer 1 (high $q$) & $\alpha =  1.005$	&	$\beta = 0.976$ & $\nu= 0.09$\\
Layer 2 (low $q$)	& $\gamma = 1.556$ &	$\delta = 0.835$ & $1-\nu=0.91$
\end{tabular}\caption{Chain deformation parameters}\label{fitparams}
\end{table}

Eq.~\eqref{fullq} is divided by its isotropic counterpart, Eq.~\eqref{debye}, and fitted in 2D to the experimental data shown in Fig.~\eqref{hiQexp}. The five fitting parameters are listed in Table~\ref{fitparams}. They were obtained by a standard genetic fitting algorithm. Using these values, Eq.~\eqref{fullq} is plotted in Fig.~\ref{hiQfit} and is seen to match the experimental data reasonably well. To assess the quality of the fit, we show the residuals (difference between the data and the fit) in Fig.~\ref{residuals}. Admittedly, some structure remains unfitted, mostly the low $q_x$ area with a difference of $0.05$. In comparison, the amplitude of the signal change in the same area is $0.3$, meaning that our fit accounts for at least $0.25/0.3=\SI{85}{\percent}$ of the observed phenomenon. The remainder is likely to be a combination of some structure factor contribution due to imperfect contrast-matching, impurities, instrument bias, and an inexact fitting function.

Using the parameters from Table~\ref{fitparams}, the piecewise model of Eq.~\eqref{regimes} is plotted in Fig.~\ref{model} with solid black lines for the $q_x$ and $q_z$ directions. Quite obviously, a realistic polymer structure cannot have sharp kinks, so we have fitted two smooth curves (dotted red and blue) to our piecewise model. These fits are made with a semi-empirical function
\begin{multline}\label{semirouse}
\frac{\braket{(\mathbf{q}\cdot \mathbf{r}_{nm})^2}}{2(qR^2)} = \left|\frac{n-m}{N}\right| +\\
\left( \frac{(B_x q_x)^2 - (B_z q_z)^2}{q^2} \right) \left|\frac{n-m}{N}\right|^{\xi} 
\end{multline}
First, the anisotropic amplitudes are fixed at $B_x = \sqrt{\nu \alpha^2 + (1-\nu) \gamma^2} = 1.51$ and $B_z = \sqrt{\nu \beta^2 + (1-\nu) \delta^2} = 0.85$, to exactly match the endpoints of the piecewise Eq.~\eqref{regimes}. Second, the exponent $\xi$ is determined by minimizing the difference between the piecewise and the smooth curves. The result is $\xi = 1.19$ and $\xi = 1.18$ for the $x$ and $z$ axes respectively. Optimizing for both axes simultaneously still leaves us with 1.19, since the $z$-axis amplitude is much smaller. This semi-empirical expression requires merely three parameters ($B_x$, $B_z$, $\xi$) to describe the entire experiment.

\subsection{Stress estimation}
SANS is a tool to measure structure on the large scale of the whole molecule. Mechanical stress, on the other hand, arises from the structure on the short scale of one molecular bond. Yet, there is considerable overlap between these two techniques, and we shall now attempt to extract the stress tensor values from our fit of the SANS data. First, SANS is measured in units of length alone, as the intensity is given in 1/cm, and the $q$-vector is in 1/nm. In contrast, stress is measured in units of Pa, or N/m${}^2$, so clearly some additional information is required to connect these two methods. Coarse-grained polymers are often described by a mechanical model of beads joined by harmonic springs, in which case the stress tensor is derived to be~\cite{mcleish2002tube}:
\begin{equation}\label{stress}
\sigma_{ij} = \left(\frac{\rho N_A k_B T}{3 M_r}\right) \frac{3\braket{r_i r_j}}{\lambda^2}
\end{equation}
where $r_i$ is the bond vector in the $i^{\text{th}}$ direction $i=(x,y,z)$. The pre-factor contains the polymer mass density, the Avogadro number, the thermal energy, and the molecular mass of a monomer. The product in the big parentheses amounts to $\sigma_0=\SI{2e6}{\pascal}$ for our polystyrene solution. To extract the bond length from SANS, we go back to Eq.~\eqref{regimes}, which says that at short distances, the polymer has the structure of an ideal random walk of step length $\braket{r_x^2} = (\alpha \lambda)^2/3$ and $\braket{r_z^2} = (\beta \lambda)^2/3$ along $x$ and $z$ respectively. Plugging this into Eq.~\eqref{stress}, we obtain a rheological quantity called the third normal stress difference
\begin{equation}
N_3 = \sigma_{xx}-\sigma_{zz} = \sigma_0 \left(\alpha^2 - \beta^2\right) = \SI{1.3e5}{\pascal}
\end{equation}
Since we could not measure this quantity with our shear apparatus, we compare it with the available literature data of a similar polymer. Ref.~\cite{kannan1992third} reports oscillatory shear results for polyisoprene of $M_w = \SI{170}{\gram\per\mole}$, which is 3 times shorter than our polystyrene, but also 3 times denser as they have used a melt instead of a semi-dilute solution. Judging from the dynamical moduli data in Fig.~3c of that study, the cross-over frequency, which corresponds to $\text{Wi}=1$, is at $\omega a_T = \SI{6e-3}{}$. To compare with our conditions of $\text{Wi}=30$, we look at their Fig.~4c and frequency $\omega a_T = 0.18$. Reading off the stress axis we find $N_3 = \SI{6e4}{\pascal}$, which is half of the magnitude that we could infer from our piecewise fit of SANS. It shows that our structural data analysis is reasonably consistent with an independent rheology perspective. We attribute the remaining discrepancy of $\mathcal{O}(2)$ partly to the difference of sample chemistry, but mostly to the uncertainty of the SANS data in the high $q$ region, which is the important bit for calculating the stress. A more precise comparison with rheology may become available in the future, by improving the resolution and the counting time of the SANS setup, and by collecting data in more directions than just the $xz$ plane.

\section{Discussion}
Our experiment can be compared to an earlier work in Ref.~\cite{Muller1993}, where an entangled PS melt has been sheared, quenched, and measured with \emph{ex situ} SANS, using the same PAXY instrument. Their data does not show any change along the vorticity axis, whereas we observe a clear increase in scattering (chain shrinkage), although the effect is $(\gamma-1)/(1-\delta) = 3.37$ times weaker than the stretching seen along the flow axis (see Table~\ref{fitparams}). This discrepancy could be explained by their slower shear rate of $\text{Wi} = 4$, compared to ours $\text{Wi} = 30$. The anisotropy in the melt case was thus entirely due to the change along the flow axis. It was quantified by fitting ellipses to the scattering data, and taking the ratio of their axes. In the flow-vorticity plane data is available for $\text{Wi} \approx 1$, where anisotropy is seen to decrease from 1.39 to 1.23, the value at which it saturates with increasing $q$. In contrast, the anisotropy in our data decreases continuously from $\gamma/\delta = 1.86$ to $\alpha/\beta = 1.03$, and is almost perfectly isotropic at high $q$. To summarize, shear experiments on \emph{ex situ} melts and \emph{in situ} semi-dilute solutions bear qualitative similarities at low $q$, but the universality breaks down at high $q$, where we see almost no saturation or plateau of the anisotropy.

We have extracted the chain deformation parameters, Table~\ref{fitparams}, using a purely structural model, without any recourse to molecular theories. Nevertheless, to understand why the chain deforms in this particular way, a molecular explanation is needed. Currently no definitive theory exists, but the main contender in this arena is GLaMM~\cite{graham2003microscopic}, a tube theory~\cite{de1971reptation} with several modifications. In essence, the many-chain fluid is simplified with just a single chain trapped in a tube, which is the mean field of other chains, and the overall dynamics are described in a self-consistent way. This model can accurately reproduce the rheology of entangled polymer melts, although SANS studies have not reached a consensus yet, with some authors claiming a strong support of tube theory~\cite{blanchard2005small}, others report no evidence of any tubes~\cite{boue1987transient}, and others still demonstrate kinetic trends opposite to theoretical predictions~\cite{wang2017fingerprinting}. The debate centers on how exactly does the tube relax, and how is it affected by a strong deformation.

There is considerable universality between entangled polymer melts and semi-dilute solutions, especially in the linear regime $\text{Wi}<1$, where GLaMM could be applied. At higher shear the universality breaks down, as polymer solutions display enhanced concentration fluctuations, which can reach length scales considerably larger than the molecule radius of gyration~\cite{helfand1989large, mendes1991experimental, hashimoto1992butterfly, boue1994semi, saito2002structures}. Therefore, a single average chain in a tube may not be enough to describe the whole fluid. Furthermore, the shape of an individual molecule is known to fluctuate between highly stretched and collapsed states, a phenomenon called tumbling dynamics~\cite{teixeira2005shear}. The mean field assumption, a core tenet of tube theory, becomes questionable given such inhomogeneities. Finally, we note that the form factor measured by SANS is a fundamentally static quantity (contains no units of time), and could be consistent with many different dynamical theories. Given the above limitations, it would be premature to interpret our findings in terms of the current tube theories.

Instead, we offer an explanation based on the fact that an entangled polymer liquid is an intermediate case between a rubber and an ideal Rouse chain. A piece of rubber responds to stress with an affine deformation, meaning that the exponent in Eq.~\eqref{semirouse} is $\xi = 1$. It is widely believed that entangled polymers, at a large scale, have this rubber-like affine response~\cite{rubinstein1997nonaffine, basu2011nonaffine}. However, on the short scale, a non-affine liquid-like response is expected. In this regime, unentangled polymers are well described by the Rouse model, which contains the following forces: spring, random, and shear (see Chapter~4 in Ref.~\cite{doi1988theory} for details):
\begin{equation}
\frac{\partial \mathbf{X}_p}{\partial t} = -\frac{k_p}{\zeta_p} \mathbf{X}_p + \frac{\mathbf{f}_p}{\zeta_p} + \kappa (\jhat \cdot \mathbf{X}_p) \ihat
\end{equation}
Its solution gives the mean square value of the Rouse modes $\mathbf{X}_p(t)$ in the thermodynamic limit $t\rightarrow \infty$:
\begin{equation}
\braket{(\mathbf{q} \cdot \mathbf{X}_p)^2} = \frac{k_B T}{k_p} \left[q^2+q_x^2\frac{(\kappa \zeta_p/k_p)^2}{2}\right]
\end{equation}
in agreement with Ref.~\cite{pincus1976excluded}. We use standard definitions for the mode friction $\zeta_p = 2N\zeta$ and the mode stiffness $k_p = 6\pi^2 k_B T p^2/(N\lambda^2)$. The above equation shows that the mean square width of a harmonic dumbbell is elongated by a factor of $1+(\kappa \tau_p)^2/2$, where $\tau_p = \zeta_p/k_p$ is its thermal relaxation time. The quadratic dependence on the dimensionless shear rate $(\kappa \tau_p)^2 = \text{Wi}^2$ is expected to hold even in the complete multi-chain theory, since it is the first non-zero term in the Taylor expansion of any reasonably behaved function, and is sufficient to describe the effect as long as the shear is not too strong. In the future a much stronger shear may become accessible, in which case we would simply argue for adding a $\text{Wi}^4$ term. The odd terms are all zero, because reversing the flow $\text{Wi} \rightarrow -\text{Wi}$ is equivalent to flipping the axis $q_x \rightarrow -q_x$, which does not affect the physics.

The pairwise distance for polymers is obtained by summing all the Rouse modes (dumbbells):
\begin{equation}
\mathbf{r}_{nm} = 2\sum_{p=1}^{\infty} \mathbf{X}_p [\cos (p\pi n/N) - \cos (p\pi m/N)]
\end{equation}
and this leads to the chain structure
\begin{multline}\label{fseries}
\frac{\braket{(\mathbf{q}\cdot \mathbf{r}_{nm})^2}}{2(qR^2)} = \frac{|n-m|}{N} +\\
\left(\frac{q_x \kappa \tau_1}{\pi q}\right)^2 \sum_{p=1}^{\infty} \frac{\braket{[\cos(p\pi n/N)-\cos(p\pi m/N)]^2}}{p^6}\\
= \mu + \frac{\pi^4}{180} \left(\frac{q_x \kappa \tau_1 \mu}{q}\right)^2 \left(1+\mu-(\mu/2)^2-2\mu^3 + \mu^4\right)
\end{multline}
where $\mu = |n-m|/N$ has been defined for brevity. The above equation is the exact analytical solution of the Rouse chain structure under shear, and is reported here for the first time. The summation of the Fourier series, Eq.~\eqref{fseries}, can be performed using formulas tabulated in Ref.~\cite{gradshteyn2014table}. We can see that for small separations $\mu\ll 1$ where the Rouse model should have some validity, it predicts the deformation exponent $\xi = 2$. Hence, our fitted value of $\xi=1.19$ lies between the rubber (1) and the liquid (2) predictions, a reasonable outcome for a viscoelastic material.

\section{Conclusion}
In this study we have performed the first SANS experiment on the form factor of entangled semi-dilute polymers under \emph{in situ} shear. We have verified our quiescent result against an interpolation of literature measurements in different solvents. We have then compared our shear result with earlier \emph{ex situ} data and found a qualitative agreement. To allow a deeper analysis, we have derived an analytical fitting function for SANS in 3D, which is the major novelty of this work. From the fit we have shown that the molecular deformation follows a power law between an elastic rubber and a Rouse liquid. In addition, we have used our SANS fit to calculate the rheological third normal stress difference, and compared the outcome with the literature data. The match was reasonably good, which is encouraging for future studies, as it is now possible to directly connect the SANS spectra with mechanical stress, in both shear and normal components. Our fit is independent of molecular theories, and is therefore applicable to deformed polymers in a wide variety of situations: dilute, semi-dilute, and melts, as well as cross-linked materials like rubbers and gels, in addition to polymer-nanoparticle composites. For such complex materials, far from equilibrium, reliable theories and simulations are yet to be developed. Traditionally, one has to postulate (guess) a theory, calculate the resulting SANS, and compare it with experiment, until a good match is found. Our piecewise fit takes the guesswork out of the equation, and instead directly provides the real-space structure, from which a theory is more straightforward to deduce.

\section{Conflics of interest}
There are no conflicts to declare.

\section{Acknoledgements}
The SANS beamtime was provided by the Laboratoire L{\'e}on Brillouin, France, instrument scientist Alain Lapp. A.K. acknowledges the financial support of the Swedish research council and the Carl Tryggers stiftelse, grant CTS 16:519. 

\section{Author contributions}
A.K. has fitted the data and has written the article. S.P. has reduced the raw data with contributions of A.D. F.A.A., P.G. and A.K. have prepared the polymer solution. M.W. has contributed to the shear cell design. M.K. has built the shear cell and has led the experiment. All authors have participated in the neutron experiments.

%
%
%


\bibliography{manuscript}

\end{document}